# Analog Optical Transmission of Fast Photomultiplier Pulses Over Distances of 2 km

A. Karle, T. Mikolajski, S. Cichos, S. Hundertmark, D. Pandel, C. Spiering, O. Streicher, T. Thon, C. Wiebusch, R. Wischnewski

*DESY, Institute for High Energy Physics, Platanenallee 6, D 15738 Zeuthen, Germany, E-mail: Karle@ifh.de*

New LED-transmitters have been used to develop a new method of fast analog transmission of PMT pulses over large distances. The transmitters, consisting basically of InGaAsP LEDs with the maximum emission of light at 1300 nm, allow the transmission of fast photomultiplier pulses over distances of more than 2 km. The shape of the photomultiplier pulses is maintained, with an attenuation less than 1 dB/km. Typical applications of analog optical signal transmission are surface air shower detectors and underwater/ice neutrino experiments, which measure fast Cherenkov or scintillator pulses at large detector distances to the central DAQ system.

## 1 Introduction

Cosmic ray surface air shower experiments consist of hundreds to thousands of detector stations spread over distances of 100 m to several km [1,2]. Underwater/ice neutrino detectors, such as AMANDA, transmit photomultiplier pulses over a distance of 2 km to a surface or shore station [3]. In such experiments fast signals are transmitted to a central data acquisition system. Until now, experimentalists usually had the choice between digital data transmission or analog transmission with the use of electrical cables. In this contribution we present an alternative method to transmit fast analog data via optical fibers. We started this development for the transmission of PMT pulses in the AMANDA experiment. In this case, the required distance is about 2 km.

Contribution to the First International Conference on New Developments in Photodetection, Beaune 96, June 1996, to appear in Proc. Suppl. Nucl. Inst. and Meth., Section A
Preprint: DESY 96-186

Table 1
Typical characteristics of coaxial cables and multimode optical fibers

| Type of cable | Coaxial Cable (RG 58 C/U) | Optical multimode @ 1300 nm (62.5/125 $\mu$m) |
|---|---|---|
| Attenuation | 174 dB/km @ 100 MHz | 10%/km @ 500 MHz |
| Weight | 36 kg/km | 1-8 kg/km * |
| Diameter | 4.95 mm | 0.25 mm ** |
| Cross talk | possible | none |

* Typical values. The weight depends on the mechanical construction and on the number of fibers used in one bundle.
** The diameter of the bare fiber. Typical jackets are 0.9 mm, and an outer jacket is typically 2.5 mm.

Table 1 summarizes important properties of coaxial cables and typical optical multimode fibers. It is obvious that the transmission properties of optical fibers are far superior to those of electrical cables. However, only recently has a new type of inexpensive and fast transmitters become available. The new LED-transmitters and PIN-photodiodes operating at a wavelength of 1300 nm allow for a straight forward and robust method of analog transmission.

## 2 The Set-up of the Analog Optical Transmission

In figure 1 the general set-up of the analog optical fiber transmission of PMT pulses is shown.

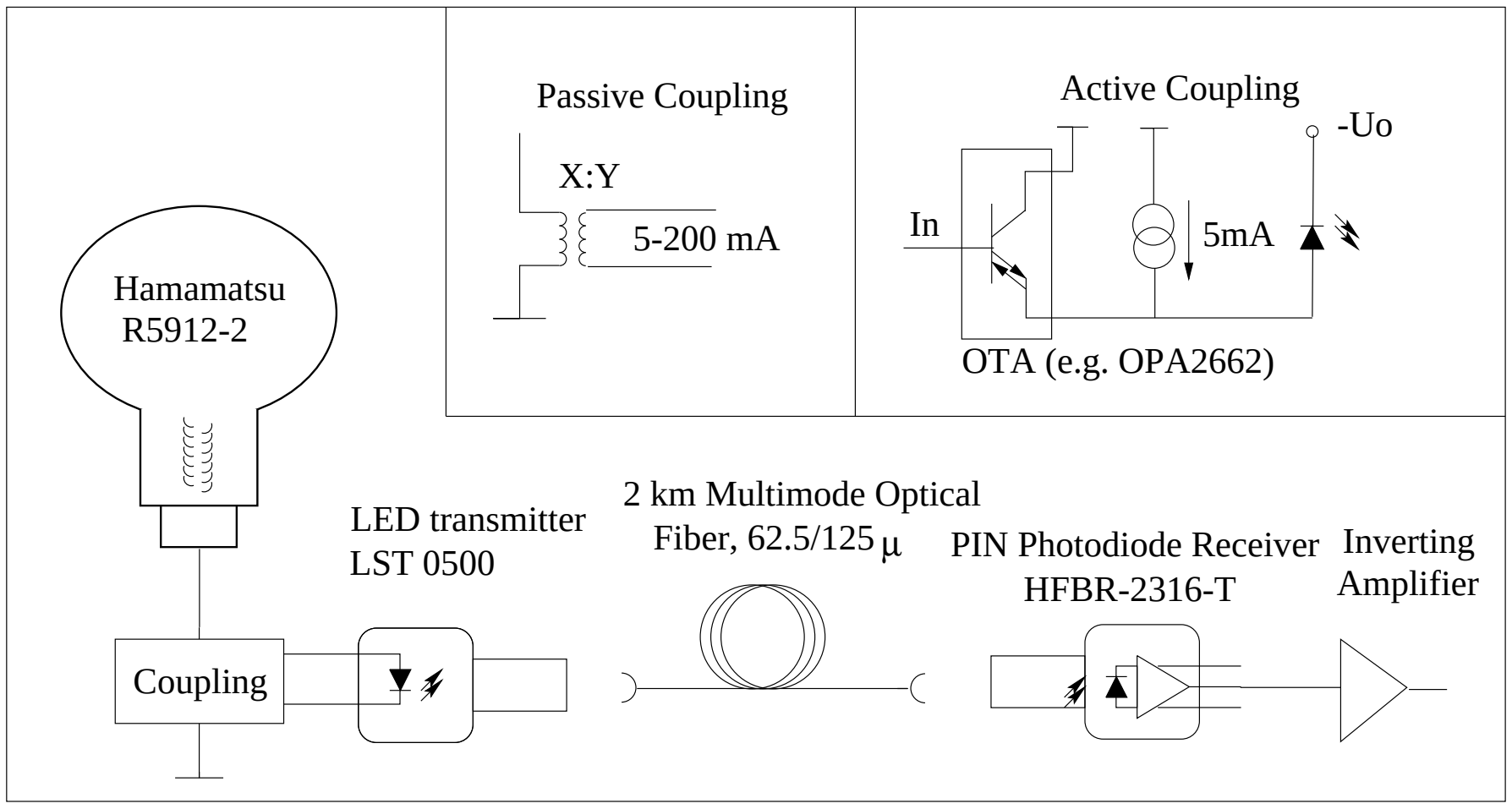

Fig. 1. Scheme for the analog pulse transmission with optical fibers.

A current pulse injected into an LED-transmitter is converted into a 1300 nm light pulse. The transmitter is an InGaAsP LED (LST-0500, Hewlett Packard) designed for fiber applications. The bandwidth of the transmitter is 255 MBd. The transmitter is connected to the receiver by a multimode optical fiber (62.5/125 $\mu$m) with a length of 2 km. The receiver is an InGaAs PIN photodiode (HFBR-2316T) with an integrated low-noise transimpedance preamplifier. Both operate at a wavelength of 1300 nm in the lower dispersion and attenuation region of standard fibers. The risetime and the FWHM of a short nsec pulse is 4 nsec after transmission over 2 km, which is consistent with the bandwidth of the receiver (125 MBd).

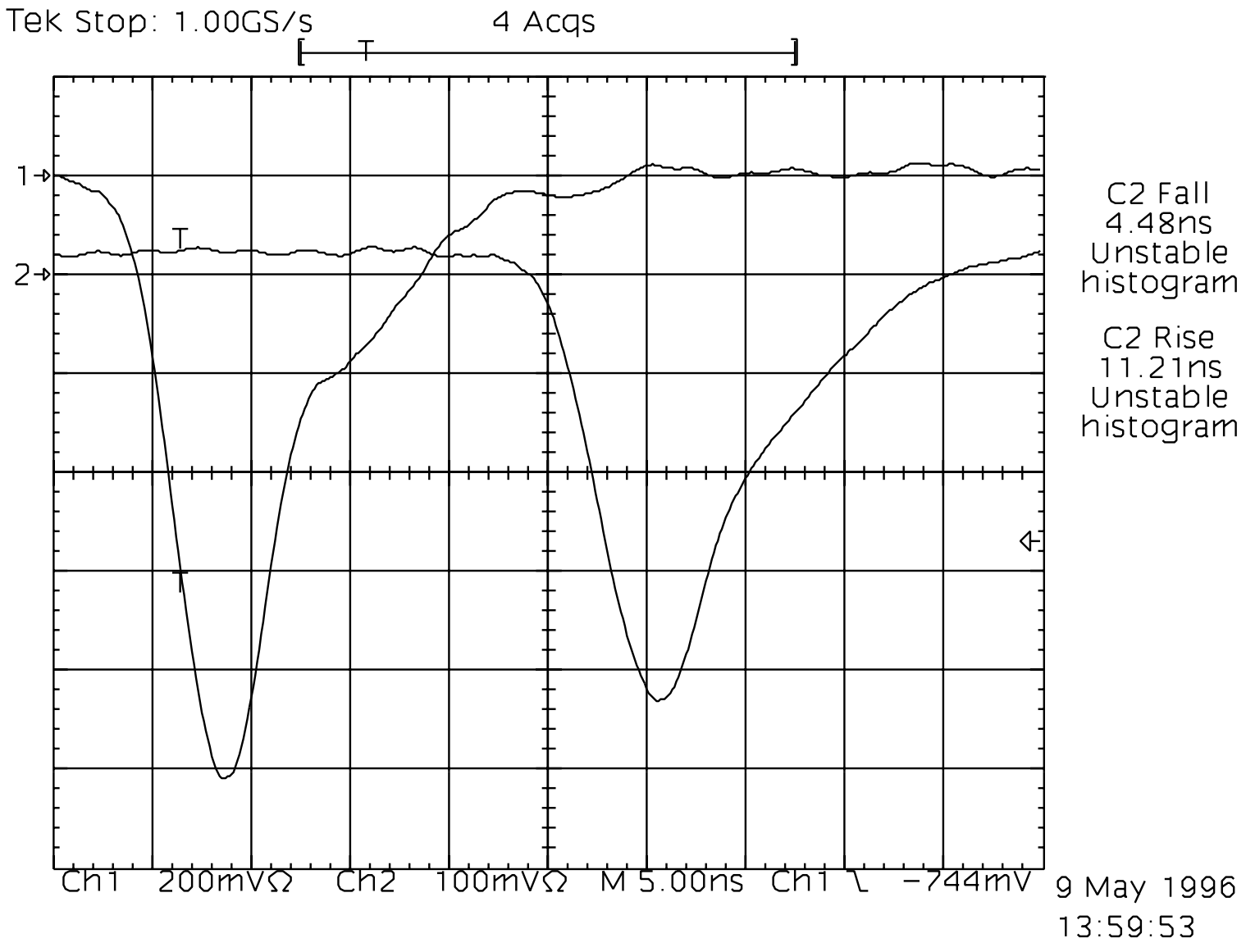

Fig. 2. An oscilloscope hard-copy of a single photoelectron PMT-pulse (left) and the same pulse after transmission over 2 km multimode fiber (right). The full range of the display corresponds to 50 nsecs.

The LED-transmitter shows linear light output from about 1 mA to 150 mA. Many photomultipliers produce current pulses in the range of a few mA to more than 100 mA. In our case, where the gain of the R5912-2 photomultiplier can be as high as $10^9$, a single PE pulse is amplified to a peak current of 20 mA. Therefore, amplifiers are not necessarily required for the coupling of photomultiplier pulses to the LED-transmitter. We tested two methods of coupling a PMT pulse into the LED-transmitter (see figure 1). In one case the PMT-current is coupled directly to the LED. A transformer can be used to modify the pulse height. With this method transformer is used as a passive current amplifier. In the case that the PMT cathode is at ground the transformer may also be useful to seperate the LED circuit from the high tension of the PMT at the anode level.

However, to obtain a high dynamic range and a good pedestal calibration it is necessary to operate the LED with a current bias of at least 1 mA. The LED should be biased since the light output of the LED becomes nonlinear if the current drops below 1 mA. Therefore, the dynamic range of the system can be improved significantly by using active coupling with a current bias in the range of 3 to 10 mA. The following measurements were done with an active coupling and a current bias of 10 mA. In an experimental environment this allows also the monitoring of the DC optical power with a conventional optical power meter.

## 3 Measurements

Figure 2 shows a single photoelectron pulse of the 14-stage photomultiplier Hamamatsu R5912-2 before and after transmission. The transmitted pulse shape is essentially identical to the original pulse shape. The achievable dynamic range in this configuration is ultimately limited by the maximum pulse height of the transmitter and the receiver (2 V), and the noise level of the receiver amplifier (1-2 mV in case of the HFBR-2316T).

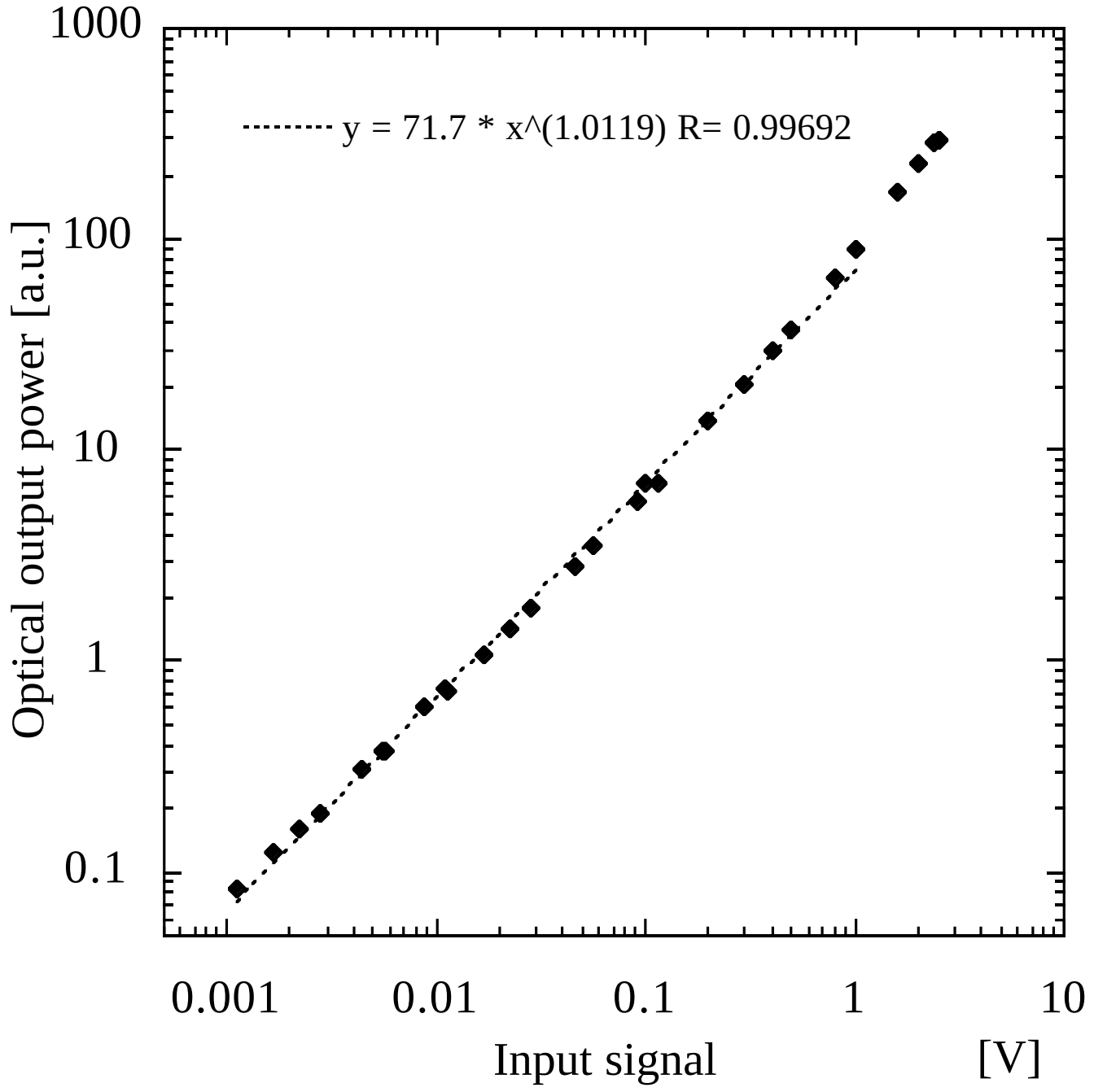

Fig. 3. Linearity of the LED-transmitter operated with active coupling and biased with 10 mA. A power law fit shows good linearity from 1 to 1000 mV.

Figure 3 shows the linearity of the LED optical output power. An input pulse of 1 mV corresponds to an LED current of approximately 130 $\mu$A. The LED-transmitter shows very good linearity over about 3 orders of magnitude. Only when pulses are larger than 1 V, does the light output deviate slightly from linearity. The maximum current through the LED corresponds to 250 mA. It should also be mentioned, that for large pulses the used receiver amplifier shows nonlinear behavior. To measure the maximum linearity over the full dynamic range, we split the optical pulse and attenuated one channel.

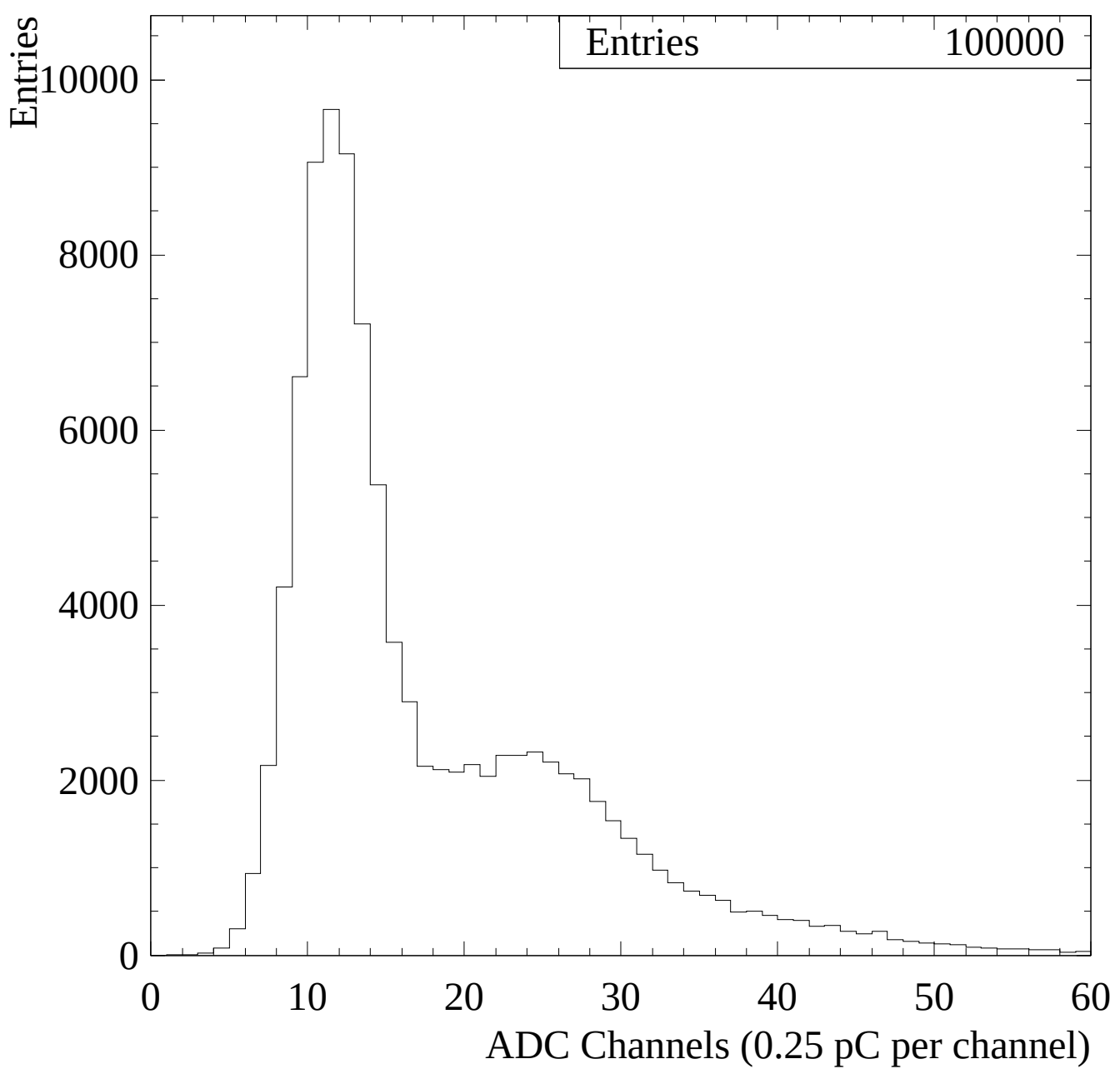

Fig. 4. Single photoelectron spectrum measured with a charge sensitive ADC. The PMT-pulses were attenuated electrically by 30 dB.

If single photoelectron resolution is required, we define the dynamic range by the maximum measurable number of photoelectrons generated by a light pulse of a duration which is shorter than 10 nsec. In this case, the smallest signal to resolve is a single photoelectron pulse. In figure 4 the charge distribution of single photoelectron pulses is shown after 2 km fiber transmission. The PMT signal was attenuated by 30 dB. From this we conclude that the achievable linear dynamic range is at least 32 photoelectrons. We expect that values larger than 100 photoelectrons can be achieved.

## 4 Applications in Cosmic Ray Experiments

This technology could be applied to several types of cosmic ray experiments, in which fast analog data must be transmitted to a central DAQ station. There the signal processing and triggering can be accomplished with standard electronics used in high energy physics. Possible applications are:

- Underwater/ice neutrino experiments: In the case of AMANDA [3,4] the photomultipliers are located about 2 km below the ice surface. It is planned to test this technology in a prototype optical module in the AMANDA experiment.

- Imaging Cherenkov telescopes [5,2]: The analog signals of imaging cameras of more than 500 pixels can be transmitted without losses in bandwidth or amplitude, and without crosstalk to a nearby laboratory. The low weight of the fiber bundle would not disturb the camera assembly.
- Surface Cosmic Ray Experiments [1,2]: Scintillator arrays, as well as wide angle Cherenkov detectors, such as AIROBICC [6], transmit analog pulses over distances larger than 100 m. Future km-scale detectors could avoid a complete digitization at the local detector station.

Due to its importance for communication networks, the technology of optical fiber transmission at the wavelength of 1300 nm is rapidly developing. New components, which should improve the dynamic range, the linearity, and the bandwidth of this technology are already being offered.